\documentclass[twocolumn]{aastex631}
\usepackage[utf8]{inputenc}
\usepackage{mathtools}
\usepackage{comment}
\usepackage{enumitem}
\submitjournal{ApJ}
\accepted{Jan 2024}
\shorttitle{Segregation and Collisions}
\shortauthors{Balberg}

\newcommand{\MBH}{M_{\bullet}}
\newcommand{\Msun}{M_{\odot}}
\newcommand{\Rsun}{R_{\odot}}

\newcommand{\e}{\varepsilon}

\newcommand{\ttb}{T_{\rm 2B}}
\newcommand{\ttbj}{T^J_{\rm 2B}}
\newcommand{\tDFj}{T^J_{\rm DF}}

\newcommand{\invyr}{yr^{\rm {-1}}}

\newcommand{\mmax}{m_{\rm max}}

\submitjournal{\apj}

\begin{document}
\title{Segregation and Collisions in Galactic Nuclei: Rates of Destructive Events Near a Supermassive Black Hole}

\correspondingauthor{Shmuel Balberg}
\email{shmuel.balberg@mail.huji.ac.il}

\author[0000-0002-0349-1875]{Shmuel Balberg}
\affiliation{Racah Institute, Hebrew University of Jerusalem, Jerusalem, 91904, Israel}


\begin{abstract}
The centers of galaxies host a supermassive black hole surrounded by a dense stellar cluster. The cluster is expected to develop mass segregation, in which gravitational scatterings among the stars cause heavier objects to sink closer to the central black hole, while lighter objects will tend to be overconcentrated in the outer regions. This work focuses on the implications of mass segregation on the different channels for violent destruction of stars in the cluster: tidal disruptions, gravitational-wave-driven inspirals and high-velocity destructive collisions between stars. All such events occur close to the central black hole, where the heavier objects congregate. The analysis is based on a simplified Monte Carlo simulation, which evolves a two-mass population in a cluster surrounding a Milky Way-like super massive black hole. The simulation is based on the single-mass scheme used by \citet{SariFragione2019} and \citet{BalbergYassur2023}, which has been extended to allow for the dynamical friction effects typical of unequal mass populations. The effects of mass segregation on the rates of the different destruction channels are analyzed self-consistently in the overall evolution of the cluster. Also considered are stars which are injected into the cluster after being disrupted from a binary system by the supermassive black hole (SMBH). Such stars are captured in the inner regions of the cluster, and so their orbital evolution, as well as their destruction rate, are therefore influenced by heavy objects that might be abundant in the vicinity of the SMBH.
\end{abstract}


\section{Introduction\label{sec:Intro}}

Supermassive black holes (SMBHs) in the centers of galaxies are typically surrounded by a dense cluster of stars. The distribution and dynamics of the stars are mostly determined by a combination of the SMBH's dominant potential and gravitational interactions between the stars \citep{Alexander2017}. These dynamics give rise to multiple channels for the violent destruction of stars in the cluster, specifically tidal disruption of stars \citep{Rees1988,Gezari2021}, gravitational-wave emission as objects inspiral into the SMBH \citep{BarackCutler2004,ASetal2007}, and high-velocity stellar collisions \citep{BalbergSariLoeb2013,AmaroSeoane2023}. While relatively rare, such events are likely to produce unique observational signatures, and reliable estimates of their rates are certainly required.

Several estimates of the rates for tidal disruption events (TDEs) and extreme-mass-ratio inspirals (EMRIs) have been carried out in the past (see, e.g., \citet{SyerUlmer1999,MagorrianTremaine1999,WangMerritt2004,BBK2011,AharonPerets2016,Panamarevetal2019,Stoneetal2020,Broggietal2022}). Less attention has been given to destructive collisions (DCs), which in fact cannot be neglected among the processes which determine the steady-state profile of the cluster. In a recent work, \citet{BalbergYassur2023} expanded the approximate Monte Carlo-based \cite{Henon1971} method, developed by \citet{FragioneSari2018} and \citet{SariFragione2019} for a single-mass population of stars, by including a model for DCs. In particular, they found that, once accounted for, DCs replace EMRIs as the secondary process which depletes the innermost part of the stellar "cusp." This is especially true when considering stars that are captured near the SMBH following binary tidal disruption - the \citet{Hills1988} mechanism. If this capture rate is significant, DCs between captured stars might occur at a rate comparable to that of TDEs.  

The single-mass population approximation mentioned above is generally considered a reasonable representation of a stellar cluster and has indeed been used in many previous studies. Specifically, the general setting can be assessed based on the groundbreaking analysis of \cite{BW76}, who developed a fundamental solution for a single-mass population of stars assumed to have an isotropic velocity field. By approximating the gravitational potential of the SMBH as Newtonian, and treating all interactions between stars as random two-body gravitational scatterings of point objects, they found that the steady state number density profile follows a power law of $n(r)\sim r^{-\alpha}$ with $\alpha=7/4$ (hereafter, the BW profile). This result, which applies when the SMBH dominates the gravitational potential, serves as a point of reference for analyses of galactic nuclei. 

In reality, of course, stars in the cluster, are  distributed in a continuous mass function. 
It is inevitable that such a stellar cluster will settle on a steady state near an SMBH that includes "mass segregation," in which heavier objects are overconcentrated at smaller distances from the SMBH and lighter objects are overconcentrated further out. This is due to the tendency toward energy equipartition in multiple gravitational encounters, which will cause the heavier objects to sink in the potential well of the SMBH. In essence, this is another manifestation of dynamical friction \citep{Chandrasekhar43}. The point of reference for mass segregation in galactic nuclei was again set by \citet{BW77}, who generalized their original work by including multiple mass groups. By limiting their analysis to the case where the massive stars are also most abundant, they found a simple analytic solution, in which the most massive stars of mass $m=\mmax$ settle on the single-mass BW profile, while lighter stars with $m<\mmax$ have a somewhat shallower steady-state profile of $n_m(r) \propto r^{-(3/2 + m/4\mmax)}$.

A sufficiently top-heavy stellar mass function is not necessarily realistic. \citet{AlexanderHopman2009} coined the term "weak segregation" to describe this case when heavier objects dominate their own scatterings all through the cluster. They demonstrated that in the opposite scenario, where heavier objects are rare and their gravitational scatterings are dominated by the lighter objects all the way to the inner boundary, there is "strong segregation" (see also \citet{PretoAS2010}). In this case the "cusp" of heavier objects settles on a density profile which is even steeper than the BW solution, with $\alpha=1.75-2.25$, depending on the details. Results for weak and strong segregation were since investigated in additional studies  \citep{KHA2009,Merritt2010,ASPreto2011,Antonini2014,AharonPerets2016,BASS2018,ZhangAS2023}, and generally reproduced the analysis by \citet{AlexanderHopman2009} (see, however, the recent critique by \cite{LinialSari2022} and below in the results of \S \ref{sec:Simple}).

Since they occur close to the SMBH, rare destructive events may be particularly sensitive to the existence of a component of heavy objects. While in some cases general relativistic effects and the impact of segregation on the EMRI rate were explicitly considered \citep{HopmanAlexander2006,AharonPerets2016,BarOrAlexander2016}, the possibility of stellar collisions in conjunction with mass segregation has not yet been addressed. Most notably, there is the possible impact on stars from disrupted binaries, which are generally "injected" close enough to the SMBH so that they constitute a major component of the stellar destruction rate \citep{SariFragione2019}. Moreover, their collision time scale is significantly shorter than the two-body relaxation time scale, and they dramatically boost the DC rate \citep{BalbergYassur2023}.  

The goal of the present work is to examine the potential effect of mass segregation on the expected rates of the violent destruction events of stars. Heavier objects that have segregated to the interior regions of the cluster will reduce the two-body relaxation time, and also cause the lighter stars to gain energy on average through dynamical friction. Both effects, if significant, may indeed modify the rates of destructive events. The evaluation of these rates is done by generalizing the code described in \citet{BalbergYassur2023} to a two-mass population. The code tracks two-body scatterings, GW losses and collisions self-consistently, and has been extended to include approximate treatment of the transfer of energy and angular momentum between the two components. The analysis includes the case when captured stars from disrupted binaries contribute a significant source to the main-sequence (MS) stellar density close to the SMBH. 

This paper is structured as follows. Section \ref{sec:2BDF} reviews the principal equations that describe two-body relaxation and dynamical friction in a two-mass system, and presents the approximations used later in this work. For completeness, the main features of the numerical code, presented in \citet{BalbergYassur2023}, are summarized in Section \ref{sec:MCGCcode}, noting the extensions introduced to describe dynamical friction. The steady-state density profiles of test simulations for a case with no relativistic effects, no collisions, and no captured stars from disrupted binaries are shown in Section \ref{sec:Simple}, and compared to standard segregation theory as mentioned above. Section \ref{sec:MainResults} presents the main results of this work, when including GW losses, DCs, and possibly a significant source of stars from binary disruptions. The presentation focuses on the corresponding rates of destructive events. Conclusions and a summary are offered in Section \ref{sec:Conclusions}.

\section{Relaxation and Dynamical Friction in a Two-Mass System}\label{sec:2BDF}

 The principal features of the stellar cluster surrounding the SMBH can be assessed based on relatively simple models \citep{BinneyTremaine2008,Merritt2013}. If two-body gravitational scatterings are completely random, they cause stars in the stellar cluster to diffuse in angular momentum and energy space. The typical time scale for a star orbiting the SMBH on a circular orbit with a semimajor-axis (sma) $r$ to change its specific angular momentum, $J_c(r)$, by order unity is
\citep{LightmanShapiro1977}
\begin{equation}\label{eq:T2B}
\begin{split}
    \ttb(r) \approx \frac{v^3(r)}{G^2 n(r)\langle m^2(r) \rangle\ln{\Lambda}} 
    \\
    \approx \frac{P(r) \MBH^2}{\left(N_H(r) m^2_L+N_H(r) m^2_H\right) \ln{\Lambda}}\,.
    \end{split}
\end{equation} 
In the first equation $n(r)$ and $v(r)$ are the stellar number density and typical velocity at $r$, respectively, and $\ln{\Lambda}$ is the appropriate Coulomb logarithm. The quantity $\langle m(r)\rangle $ is the average mass of the stars in the vicinity of $r$. In the second equation it is assumed that the stellar cluster is a two-mass population with masses $m_L$ and $m_H$ for the lighter and heavier stars, respectively (and using the approximate relation $n(r)\sim (N_L(r)+N_H(r))/r^3$). Other aspects of the second equation reflect a simplifying assumption that the SMBH mass, $\MBH$, completely dominates the gravitational potential. This should apply inside of the radius of influence, $R_h$, defined as the radial distance at which the enclosed stellar mass is equal to $\MBH$. Substituting appropriately that $v(r)\sim (G\MBH/r)^{1/2}$, and $P(r)\sim (r^3/G\MBH)^{1/2}$ is the orbital period of the star yields the final result of the right-hand side. In the case of a single-mass population equation \ref{eq:T2B} reduces to a simple expression with $N(r)m^2$ in the denominator. Note that $N_L(r)$ and $N_H(r)$ represent the number stars of each type in the vicinity of $r$, which can effectively scatter a test star at $r$. These values roughly correspond to the total number of stars of each type enclosed up to a radius $r$ (as long the number density profile is not $n(r)\sim r^{-4}$ or steeper, which is a reasonable assumption in all that follows). 

The energy relaxation time is similar (slightly longer), and changes modestly even for more eccentric orbits with the same sma $r$. However, the angular momentum relaxation time depends on the eccentricity of the orbit. The diffusive (quadratic) time evolution of angular momentum with time implies that the time scale for an order-of-unity change in the specific angular momentum of an eccentric orbit is smaller by a factor of $(J(r,r_p)/J_c(r))^2=r_p/r$, where $r_p$ is the periapse distance. The relevant two-body time scale eccentric orbits is therefore \citep{BinneyTremaine2008},  
\begin{equation}\label{eq:T2BJ}
\ttbj(r,r_p)\approx\frac{P(r) \MBH^2}{\left(N_L(r)m^2_L+N_H(r)m^2_H\right)\ln {\Lambda}}\left(\frac{r_p}{r}\right) \;.
\end{equation}

The relaxation times in equations \ref{eq:T2B} and \ref{eq:T2BJ} are independent of the mass of the test star. This simply reflects the fact that gravitational acceleration depends only on the scattering stars, which are represented by the second moment of the mass distribution. Equation \ref{eq:T2BJ} thus provides a first indication regarding the possible influence of segregation on the DC rate.  When stars captured in eccentric orbits following binary disruptions are the dominating source for collisions, both the rate and the spatial distribution are influenced by the fact that these stars are captured on very eccentric orbits. Specifically, \citet{BalbergYassur2023} found that for the Milky Way SMBH, if stars are captured near the SMBH at a rate of $10^{-5}\;\invyr$, equilibrium created by collisions is characterized by a steady-state number of a few thousand captured stars. Given the quadratic dependence on mass in the relaxation time, even a small number of heavy objects which have segregated to the inner part of the cluster can shorten the relaxation time, and possibly modify the DC rate and distribution. 

The two-body relaxation time is derived by assuming that the effect of gravitational scatterings is limited to a random-walk diffusion in angular momentum space. Any specific test star will also experience a net drift in angular momentum and kinetic energy due to dynamical friction created by the other stars. In general, the typical rates of drift in the velocity in the direction of motion, $v_\|$, and the specific kinetic energy, $E$, of a single test star of mass $m_t$ in a field of stars with mass $m_f$ are
\citep{BinneyTremaine2008,Merritt2013}
\begin{equation}\label{eq:D_vparDF}
\frac{dv_\|}{dt}\approx -\frac{G^2 m_f(m_t+m_f)\ln{\Lambda}}{v^2_t}\int_0^{v_t} dv_f v^2_f f(v_f)\;,
\end{equation}
and
\begin{equation}\label{eq:D_EDF}
\begin{split}
\frac{dE}{dt}\approx G^2 m_f\ln{\Lambda}\times\\
\left[m_t\int_0^{v_t} dv_f \frac{v^2_f}{v_t} f(v_f)-m_f\int_{v_t}^\infty dv_f v_f f(v_f)\right]\;
\end{split}
\end{equation}
(numerical prefactors of $16 \pi^2$ have been dropped). In equations \ref{eq:D_vparDF} and \ref{eq:D_EDF}, $v_t$ is the velocity of the test star, and $f(v_f)$ is the velocity distribution function of the field stars, assumed to be uniform and isotropic.

Unlike random scatterings, these dynamical friction formulae are linear and also depend on the mass of the test star, as it creates a wake in the field stars and friction is the result of the back reaction on the test star itself. The sign in equation \ref{eq:D_EDF} conforms with the standard of setting the energy of the bound stars as positive. 

Equations \ref{eq:D_vparDF} and \ref{eq:D_EDF} explicitly reflect the standard approximation that the net specific contribution to dynamical friction of a field star depends on whether its velocity is lower or higher than $v_t$. Specifically, in equation \ref{eq:D_EDF} it is assumed that field stars with velocity $v_f<v_t$ combine to affect a cooling term, while stars with $v_f>v_t$ induce a heating term. Henceforth, it is assumed that at any radial distance from the SMBH $r_x$, all stars have a velocity distribution function which is tight, symmetric, and centered around $v(r_x)=(GM/r_x)^{1/2}$. The rates of change of the parallel component of the velocity and the specific energy in a circular orbit are correspondingly simplified to 
\begin{equation}\label{eq:D_JDF_app}
\frac{dJ}{dt}\approx r\frac{dv_\|}{dt}\approx\pm \frac{J}{v}\frac{G^2 m_f(m_t+m_f)}{v^2(r)} \ln{\Lambda}\frac{1}{2}n_f(r)\;,
\end{equation}
and
\begin{equation}\label{eq:D_EDF_app}
\frac{dE}{dt}\approx \mp \frac{G^2 m_f
(m_t-m_f)}{v(r)}\ln{\Lambda}\frac{1}{2}n_f(r)\;,
\end{equation}
where $n_f(r)$ is the number density of field stars  at $r$. The factor $1/2$ reflects the assumption that approximately half of the field stars have velocities $v_f<v_t$. 

In general, dynamical friction will occur in any combinations of masses, including a single-mass case $(m_t=m_f)$. However, given that the focus of this work is on mass segregation which arises in a two-mass system $(m_t\neq m_f)$, only net transfer of angular momentum and energy between the two populations is considered. It is also naturally accessible within the approximations used in the Monte Carlo code described below (as opposed to dynamical friction within a single-mass population, which requires an actual velocity distribution). This constraint arises directly for the rate of change of the energy (equation \ref{eq:D_EDF_app} which is identically zero for $m_t=m_f$), but the 
approximate form for transfer of angular momentum (equation \ref{eq:D_JDF_app}) is also only applicable for the two-mass case. Correspondingly, the signs of changes in equations \ref{eq:D_JDF_app} and \ref{eq:D_EDF_app} are set so the top (bottom) 
option applies to the lighter (heavier) star. This choice leads to dynamical friction creating the required bias in the diffusive nature of gravitational scatterings. It effects a deterministic gain in angular momentum and loss in energy for the lighter stars, and conversely for the heavier stars, again - maintaining the convention that bound orbits have positive energies.

The approximate forms of equations \ref{eq:D_JDF_app} and \ref{eq:D_EDF_app} obviously do not contain a full description of dynamical friction processes. Specifically, \cite{AntoniniMerritt2012} demonstrated that this approximation is not truly appropriate for stars orbiting an SMBH (as opposed to a Maxwellian distribution, for which \citet{Chandrasekhar43} introduced the approximation). In fact, even light stars with $v_f>v_t$ will on average generate some dynamical friction on the heavier star's velocity. Nonetheless, given the level of the approximation used in the present work, equations  \ref{eq:D_JDF_app} and \ref{eq:D_EDF_app} will be used as an approximate description of dynamical friction among the stars, which ultimately leads to mass segregation.

Dynamical friction sets another gravitational time scale in the system. The time scale for a change of order unity in the angular momentum of a star on a roughly circular orbit through the drift effect can be estimated from equation \ref{eq:D_JDF_app} to be 
\begin{equation}\label{eq:T_DF}
\begin{split}
\tDFj(r;L/H) & \approx\frac{v}{|dv_\|/dt|}\\ & \approx \frac{P(r)\MBH^2}{N_{H/L}(r)m_{H/L}(m_H+m_L)\ln{\Lambda}}\;.
\end{split}
\end{equation}

Given that the rate of change of specific angular momentum is $J^{-1}dJ/t=v^{-1}dv_\|/dt$ regardless of eccentricity, equation \ref{eq:T_DF} actually applies to all orbits, independent of the ratio $r_p/r$. Thus, equation \ref{eq:T_DF} shows the general dynamical friction time, $\tDFj(r,r_p;L/H)$ of a star with sma $r$ and periapse $r_p$.
Note the distinction between the lighter ($L$) and heavier ($H$) stars, explicitly introducing a dependence on the mass of the scattered star. Generally, we will find that $T_{DF}(r,r_p;L)\neq T_{DF}(r,r_p;H)$. Moreover, the ratio $\tDFj(r,r_p;L/H)/\ttbj(r,r_p)$ indicates which of the two processes is more efficient in terms of driving the evolution of a particular component of the population for a given combination of $(r,r_p)$. For the heavier stars,
\begin{equation}
\label{eq:Tratio}
\frac{\tDFj(r,r_p;H)}{\ttbj(r,r_p)}\approx \frac{N_L(r)m^2_L+N_H(r)m^2_H}{N_L(r)m_L (m_L+m_H)} \left(\frac{r_p}{r}\right)^{-1}\;.
\end{equation}

The dynamical friction time scale can be either longer or shorter than the diffusive two-body relaxation time scale, depending on the ratio $m_L/m_H$ and on the local ratio of $N_L(r)/N_H(r)$. In fact, the dominant mechanism (swing or drift) may be different for the lighter and heavier stars. It is noteworthy that drift is most relevant for the heavier stars in the strong segregation limit where $N_H(r) m^2_H << N_L(r) m^2_L$, so that the dynamical friction time scale for the heavier stars on circular orbits is shorter by a factor of $m_L/m_H$. In the opposite weak segregation \citep{BW77} case in which the heavier stars dominate, $N_H(r) m_H > N_L(r) m_L$, the dynamical friction time scale may also be the shortest for lighter stars, but only by a moderate factor of $(1+m_L/m_H)^{-1}$. 

\section{Monte Carlo Simulation with Two Masses and Dynamical Friction}\label{sec:MCGCcode}

Stellar clusters surrounding an SMBH are often analyzed by Fokker-Planck solutions of diffusion in energy and angular momentum, specifically in a reduced one-dimensional formulation by assuming an isotropic velocity field \citep{BinneyTremaine2008,Merritt2013}. However, studying specific complex phenomena related to the stellar cluster requires more comprehensive solutions for the properties of the cluster. Exact solutions are currently unavailable: First and foremost, because of the need to track a very large number of objects over an extreme range of time scales. Correspondingly, many approximate methods have been suggested, which vary in content and technique depending on the particular applications. 

The cluster evolution numerical code used in this work is an extension of the single-mass code presented in  \citet{BalbergYassur2023}. In the following, the general key features and approximations used are summarized, and the main focus of the section is on the extension to a two-mass population of stars: The implementation in two-body random scatterings and the introduction of dynamical friction (\S  \ref{subsec:eqsfor2BDF}). Other features of the code are described briefly in \S \ref{subsec:MCGCother} and the reader is referred to \citet{BalbergYassur2023} for further detail.  

\subsection{Monte Carlo Code Setup} \label{subsec:MCGCsetup}

The simulations were carried out with a simplified Monte Carlo algorithm. It is based on the method originally suggested by \cite{Henon1971}, and applies the technical approach presented by \cite{FragioneSari2018} and \cite{SariFragione2019}. The code evolves a cluster of stars surrounding an SMBH, including an effective treatment of random two-body scatterings, GW losses and a procedure for allowing and tracking destructive high-velocity collisions.  

Each star is tracked individually by its specific energy, $E$, and specific angular momentum, $J$. These are translated to an sma $r$,  periapse $r_p$, eccentricity $e$ and period $P$, assuming an instantaneous Keplerian orbit in the potential of the SMBH. The specific energy is therefore $E=G\MBH/2r$. The cumulative effect of gravitational scatterings off other stars is averaged over the instantaneous stellar profile, assuming spherical symmetry.

While $r$ and $r_p$ are continuous and are stored individually for each star, it is numerically convenient to divide the $(r,r_p)$ plane into evenly spaced logarithmic bins. The intervals are set as $r_{i+1}=2r_i$, $r_{p,j+1}=2r_{p,j}$, so $i$ and $j$ serve as indices of the resulting two-dimensional grid. The structure of the cluster at each time step is then also represented by the number of stars of each type in every particular bin, $N_L(i,j)$ and $N_H(i,j)$. The time-dependent profile of the cluster is described by the instantaneous total number of stars with an sma enclosed in the radial bin $i$, regardless of periapse, $N_{L/H}(r_i)=\sum_j N_{L/H}(i,j)$. 

\subsection{Time step}
\label{subsec:timestep}

It is computationally prohibitive to evolve all of the stars in the cluster in every time step. Grouping the stars according to bins in the $(r,r_p)$ plane allows to selectively evolve only a subset of the stars - specifically, those for which a sufficient time has passed since their last evolutionary update \citep{FragioneSari2018,SariFragione2019}. In order to accomplish this, at the beginning of each time step the simulation scans the characteristic time scales for every bin in the $(r,r_p)$ plane. These are the two-body relaxation time scale, and the light-mass dynamical friction and heavy-mass dynamical friction time scales, respectively:
\begin{equation}\label{eq:ttbj_ij}
\Delta \ttbj(i,j)= F \left(\frac{r_i}{R_h}\right)^{1/2}\frac{r_{p,j}}{R_h}
\frac{\MBH^2}{\sum_s N_s(r_i)m^2_s}\;,
\end{equation}
\begin{equation}\label{eq:tDFj_L}
\Delta \tDFj(i,j;L)= F \left(\frac{r_i}{R_h}\right)^{3/2}\frac{\MBH^2}{N_H(r)m_H(m_L+m_H)}\;,
\end{equation}
\begin{equation}\label{eq:tDFj_H}
\Delta \tDFj(i,j;H)= F \left(\frac{r_i}{R_h}\right)^{3/2}\frac{\MBH^2}{N_L(r)m_L(m_L+m_H)}\;.
\end{equation}
where $\sum_s N_s(r_i)m^2_s=N_L(r_i)m^2_L+N_H(r_i)m^2_H$. Equations \ref{eq:ttbj_ij}-\ref{eq:tDFj_H} include a prefactor of 
\begin{equation}\label{eq:Fdef}
F=f_t\frac{P(R_h)}{\gamma ln {\Lambda}}\;,
\end{equation}
where $P(R_h)$ is the orbital period at the radius of influence, $\gamma=1.5$ is a control parameter used to account for the finite bin size \citep{FragioneSari2018}, and $f_T=0.1$ is the fraction of the relevant time scale allowed in a single evolutionary time step. The Coulumb logarithm is set as $\ln \Lambda =10$ \citep{Merritt2013}. The above equations are derived using the approximate relations $v(r_i)\approx (G \MBH/ r_i)^{1/2}$, $P(r_i)=(r^3_i/G\MBH)^{1/2}$ and $n(r_i)\approx N(r_i)/r^3_i$ (for each species separately).

At the beginning of each time step the code finds the minimum of the three time scales for each $(r,r_p)$ bin, $\Delta t_{min}(i,j)$, and then sets the global time step, $\Delta t$, according to the minimum over the entire grid:
\begin{equation}\label{eq:dt}
\Delta t=\min_{i,j} t_{min}(i,j)\;.
\end{equation}

Finally, the evolution due to gravitational scatterings is tracked with the scheme suggested by \cite{SariFragione2019}. For each grid cell $(i,j)$ the simulations keep the last time it was updated, $t_{old}(i,j)$. Stars in the grid cell are updated with gravitational scattering effects only when the current time, $t$, satisfies
\begin{equation}
\label{eq:update}
t-t_{old}(i,j)>\Delta t_{min}(i,j)\;.
\end{equation}
The time steps are typically a few tenths of $P(R_h)$, while the cluster relaxation time, $\ttb$, is of order $10^5P(R_h)$.  It is this scheme that allows to evolve what is a usually a small fraction of the stars in each time step, and still track $\sim 10^6$ stars over the entire length of the simulation.

Equations \ref{eq:ttbj_ij}-\ref{eq:tDFj_H} hold an implicit approximation regarding the number of stars which effectively scatter a given test star at its radial position $r_\star$. The equations substitute this number with the number of stars with an sma that corresponds to the $i$-th radial bin which includes $r_\star$. This is a reasonable assumption for large $N(r_\star)$, which is generally the case in the outer part of the simulation zone. On the other hand, close to the SMBH the numbers of stars are smaller, but the relative effect of two-body scattering is also small to begin with. Hence, this approximation is viable and is applied in the calculations reported below. For further validation, several calculations were tested with a different approximation in which the code calculates in every time step the actual effective number density of stars, $n(r)$, in each radial shell by finding the fraction of the orbit each star spends in every radial shell (easily found for Keplerian orbits around the SMBH, as is done for tracking DCs : see equations 28-30 in \citet{BalbergYassur2023}). The results were generally similar, but this method is inferior since it includes an inevitable inconsistency. At some radius the scheme for counting stars must be changed back from $n(r_i)r^3_i$ to $N(r_i)$, since in the region $r>0.1R_h$, where most of the stars reside, some stars spend parts of their orbits beyond $R_h$. In reality, other stars penetrate from $r>R_h$ to complete $n(r)$, but these are not included in the simulation. The qualitative and quantitative results were found to be very similar, and therefore the approximation that the effective number of scattering stars at $r_\star$ is $N(r_\star)$ is used in all simulations reported here.

\subsection{Evolving the Specific Energy and Angular Momentum by Gravitational Scatterings}\label{subsec:eqsfor2BDF}

In every time step the $(r,r_p)$ bins which need to be updated are determined according to equation \ref{eq:update}. The specific energy and angular momentum of stars which correspond to these bins are then evolved as follows. 
Each star is processed with a time $\Delta t=\Delta t_{min}(i_*,j_*)$, where $i_*$ and $j_*$ are the indices which correspond to its location in the $(r,r_p)$ grid. The inspected star travels along its orbit through a range of distances from the SMBH lying between the periapse $r_{p*}$ and the apoapse $r_{a*}$. These correspond to a range of  $[W_1,W_2]$ values in the radial grid. During a time step $\Delta t$ the inspected star crosses through the $W$-th bin $\Delta t/P_*$ times, for 
a total of $N(r_W)\Delta t/P_*$ interactions.

Following \citet{SariFragione2019}, the cumulative effect of random scatterings by stars in a given bin is estimated as follows. The typical change in the specific energy of the inspected star by a single field star with mass $m_f$ in the $W$-th bin is of order $\Delta E_1\approx Gm_f/r_W$. Given the random-walk nature of the diffusion in energy space, the cumulative effect of random scatterings by all $N_L(r_W)+N_H(r_W)$ stars in the $W$-th radial bin during a time step $\Delta t_{min}(i_*,j_*)$ is therefore a characteristic change in specific energy of
\begin{equation}\label{eq:DeltaES}
\Delta E_{RS}(W)=\frac{G}{r_W}\left(\sum_s N_s(r_W)m^2_s \frac{\Delta t_{min}(i_*,j_*)}{P_*}\right)^{1/2} \;.
\end{equation}
Here $P_{*}$ is the orbital period of the inspected star (typically shorter than the time interval $\Delta t_{min}(i_*,j_*)$, except for stars with $R_\star\approx R_h$). The square root reflects the diffusive (random-walk) nature of repeated scatterings. 

The characteristic change in angular momentum due to random scatterings with the stars in the $W-$th bin during the given time step is
\begin{equation}\label{eq:DeltaJS}
\Delta J_{RS}(W)=r_W\frac{\Delta E(W)}{v(W)}\;,
\end{equation}
where $v(W)=\sqrt{G\MBH/r_W}$ is the characteristic velocity at $r_W$. 

The effects of dynamical friction must be calculated separately for the lighter and heavier stars. The typical changes in the specific energy due to dynamical friction between the two populations are calculated per bin $W$ by applying equation (\ref{eq:D_EDF_app}) above and substituting for each species of stars in the partial number densities, and maintaining consistency with the approximate forms of the typical changes in energy and angular momentum from random scatterings at $r_W$ (equations \ref{eq:DeltaES} and \ref{eq:DeltaJS}).  Approximating that for each species $n(r_W)r^2_W v(r_W) \Delta t\approx N(r_W)\Delta t/P_*$ yields 
\begin{equation}\label{eq:DeltaEDF_L}
\begin{split}
\Delta E_{DF,L}(W)=& -\frac{1}{2}\frac{G^2}{r^2_W v^2(r_W)}(m_H-m_L) \times \\ N_H(r_W)m_H
 & \frac{\Delta t_{min}(i,j)}{P_{k*}} \;,
\end{split}
\end{equation}
and 
\begin{equation}\label{eq:DeltaEDF_H}
\begin{split}
\Delta E_{DF,H}(W)=& +\frac{1}{2}\frac{G^2}{r^2_W v^2(r_W)}(m_H-m_L) \times \\ N_L(r_W)m_L
 & \frac{\Delta t_{min}(i,j)}{P_{k*}} \;,
\end{split}
\end{equation}
for the changes in energy of the lighter and heavier stars, respectively. 
Similarly, by applying equation \ref{eq:D_JDF_app}, the changes in specific angular momentum through dynamical friction are calculated as
\begin{equation}\label{eq:DeltaJDF_L}
\begin{split}
\Delta J_{DF,L}(W)=& \frac{1}{2}\frac{G^2}{r_W v^3(r_W)}(m_H+m_L) \times \\ N_H(r_W)m_H
 & \frac{\Delta t_{min}(i,j)}{P_{k*}} \;,
\end{split}
\end{equation}
and 
\begin{equation}\label{eq:DeltaJDF_H}
\begin{split}
\Delta J_{DF,H}(W)=& -\frac{1}{2}\frac{G^2}{r_W v^3(r_W)}(m_H+m_L) \times \\ N_L(r_W)m_L
 & \frac{\Delta t_{min}(i,j)}{P_{k*}} \;.
\end{split}
\end{equation}
These choices conserve total energy and angular momentum in terms of the transfer between the two populations. Note that the changes induced by dynamical friction are linear in time, representing the deterministic drift caused by this effect.

Finally, the changes in specific energy and angular momentum of the inspected star due its passage in the $W$-th bin during the $\Delta t(i_*,j_*)$ time interval are updated by adding the effect of random two-body scatterings \citep{FragioneSari2018, SariFragione2019, BalbergYassur2023} and the deterministic drift of dynamical friction. Denoting the values at the beginning of the time step as $E_{old}$ and $J_{old}$, the new specific energy and angular momentum are
\begin{eqnarray}
\label{eq:step2BE}
E_{new}&=&E_{old}+\sin{\chi} \Delta E_{RS}+\Delta E_{DF} \\
\label{eq:step2BJ}
J_{new}&=&\left({J_{old}^2+\Delta J^2_{RS}-2 J_{old}\Delta J_{RS} \cos \Phi}\right)^{1/2} +\Delta J_{DF}\ ,
\end{eqnarray}
with $0\le \chi < 2\pi$ and $0\le \Phi < 2\pi$ drawn randomly from a uniform distribution. The star's orbital properties, $r_*$,$P_*$, $r_{p*}$, and $e_*$, are then updated consistently. The process is repeated for the entire range $[W_1,W_2]$ relevant for the inspected star, and for all stars in the $(i,j)$ bins that are being advanced in the global time step.

\subsection{Additional Features of the Monte Carlo Code} \label{subsec:MCGCother}

Other physical processes are tracked in the Monte Carlo code by the methods described in \citet{BalbergYassur2023}. 
\begin{enumerate}
\item GW losses due to the gravitational field of the SMBH are calculated separately for each star in each time step with standard formulae \citep{Peters1964,HopmanAlexander2005,SariFragione2019}.
\item A MS star, is considered lost when its periapse drops below its tidal disruption radius, $R_T$, which depends on the mass of the star $m_\star$ and its radius, $R_\star$. The standard value is \citep{Stoneetal2020}
\begin{equation}\label{eq:R_T}
R_T=\left(\MBH/m_\star\right)^{1/3}R_\star\;.
\end{equation}
If the star drops below $R_T$ following a gravitational scattering step, the star is considered as having undergone a TDE, while if this occurs after a GW loss step, the star is considered as having been disrupted gradually in a MS-EMRI. 
\item A compact object is considered lost when its periapse drops below $4R_S$ \citep{Merritt2013}, where
\begin{equation}\label{eq:R_S}
R_S=\frac{2 G\MBH}{c^2}\;
\end{equation}
is the Schwarzschild radius (and $c$ is the speed of light). In principle, compact objects are also lost in two types of evolutionary scenarios, either being scattered into eccentric orbits with $r_p<4R_S$ or by gradually losing energy in GW emission over multiple quasi-circular orbits. The former, often referred to as "direct plunges" \citep{HopmanAlexander2005,BarOrAlexander2016}, are the equivalent of TDEs for the MS stars, while the latter produce "clean" (no disruption) general relativistic EMRIs. While it would be interesting to quantify the details of the evolutionary tracks of individual compact objects (and estimate the resulting GW signal), accurate modeling of the strong relativistic effects close to $R_S$ are beyond the scope of this work, and will be addressed separately.
\item DCs are assumed to occur only between MS stars and only below a radius $R_{col}$, at which the typical gravitational energy in the orbit, $G\MBH m_\star/R_{col}$, equals the gravitational binding energy of the star, $G m^2_\star/R_\star$. For the Milky Way SMBH and Sun-like stars, $R_{col}\approx 4\times 10^{-3}R_h$. The probability a star that has a periapse $r_p\leq R_{col}$ will collide with another MS star during the time step is estimated by finding the instantaneous optical depth for the star along its orbit. The optical depth is calculated per radial bin, $i$, using the weighted probability of all other stars to be found in the $i$-th bin (derived from each star's orbital parameters). This allows to assess the overall number density of other MS stars in the bin, $n_L(r_i)$. The number density is coupled to a cross section of $\pi (f_R R_\star)^2$, with $f_R$ being a free parameter, taken here as $f_R=1$. The probability for a collision of a particular star during the current time step is then compared to a randomly drawn number, and if the star is determined to have undergone a DC, the details are completed in two additional stages. First, the radial bin of the collision is determined by drawing a second random number and comparing it to the partial probabilities for a collision in each bin. Given the radial bin where the collision occurred, the partner MS star is found by comparing a third random number to the weighted probabilities of finding each of the other stars in that radial bin.
\item Stars which were lost due to a TDE, an EMRI, or a DC are replaced with new stars, maintaining a constant number of stars in the simulation. The new star is placed in the simulation through one of two options: (i) representing a star that was scattered into the cluster from outside of the SMBH radius of influence and thus placed in the outer radial bin (between $R_h/2$ and $R_h$) or (ii) representing a star that was captured close to the SMBH following binary tidal disruption. The relative probabilities for each option are a control parameter of the simulation, and can be adjusted in order to produce a desirable value for the injection rate of stars from disrupted binaries. The properties of injected stars are discussed below in Section \ref{subsec:AddBin}.
\end{enumerate}

\section{Results for Fundamental Cases of Mass Segregation} \label{sec:Simple}

All the simulations described below were carried out assuming a Milky Way-like system with an SMBH of mass $\MBH=4\times10^6\;\Msun$, surrounded by a cluster of stars which extends up to a radius of influence of
\begin{equation}
R_h=\frac{GM}{\sigma^2}\approx 2\ \mathrm{pc}\ ,
\end{equation}
where $\sigma$ is the measured velocity dispersion external to the radius of influence \citep{MerrittFerrarese2001}. The two-mass population is divided between stars with mass $m_L=1\;\Msun$ and mass $m_H=10\;\Msun$, set up so that the total mass is $\sum m=\MBH$. 
The lighter stars are assumed to be MS, Sun-like stars with a radius $R_\star=\Rsun$, while the heavier stars are essentially stellar-mass black holes and approximated as point objects.
The MS stars occupy the range between the tidal radius $R_T\approx 1.1\times 10^{13}\;\mathrm{cm}$, and $R_h$, while the stellar black holes can sink as low as $4R_S$. This setup of a two-mass population serves as a reasonable approximation for studying the general physics of mass segregation of a multi-mass system 
\citep{AlexanderHopman2009}.

Simulations are initiated by predetermining the total abundances of each species, $N_L$ and $N_H$. At the beginning of the simulation, all stars are assigned values of their sma, $r$, and specific angular momentum, $J$. The sma is drawn randomly from a weighted distribution that produces a power-law profile $n(r)\propto r^{-\alpha}$, or $N(r)\sim r^{3-\alpha}$. 
The specific angular momentum is set by uniformly sampling $(J/J_C(r))^2$ over the range 0 to $J_C(r)=\sqrt{GMr}$ (the circular angular momentum given the star's sma, $r$). This choice generates a thermal distribution, equivalent to an isotropic velocity field. The values of $r$ and $J$ are then used to find the specific energy, $E$, period, $P$, eccentricity, $e$, and periapse, $r_p$, for each star assuming Keplerian orbits around the SMBH.
The initial profile is set as $\alpha_L=\alpha_H=7/4$ for both species of stars. This choice generally allows for convergence over time scales of a few tenths of $\ttb(R_h)$ (although in cases with $N_H\leq 4\times 10^3$ convergence is faster when starting with $\alpha_H>7/4$, as is predicted by the theoretical  results for strong segregation; \cite{AlexanderHopman2009,KHA2009}) .

\begin{figure}
    \centering
    \includegraphics[width=\columnwidth]{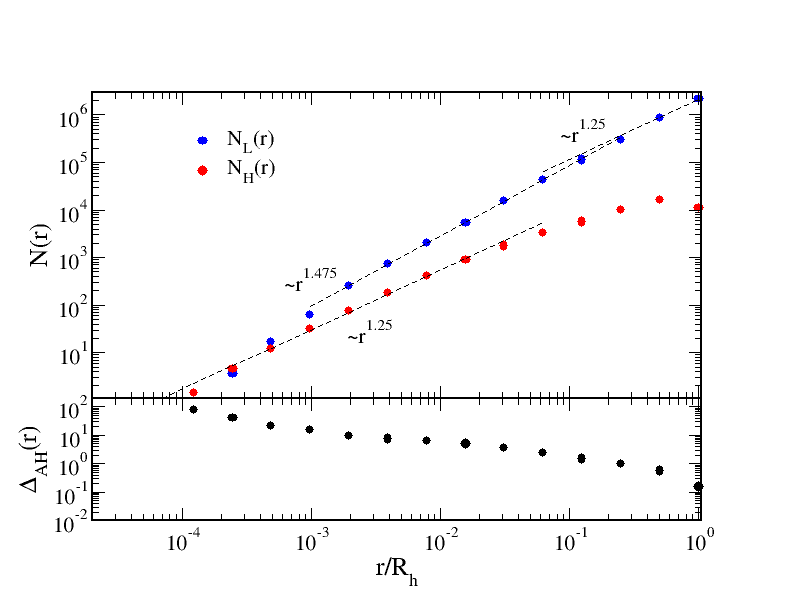}
     \caption{Top panel: the steady-state profile at time $\ttb(R_h)$ of the $m=1\Msun$ MS stars (blue) and the $m=10\Msun$ black holes (red), for the case $\{N_L=3.5\times 10^6,\;N_H=5\times 10^4\}$. GW losses, collisions, and injected stars are not included. The profile is presented through $N_L(r),\;N_H(r)$, which are the number of stars in each radial shell between $r/2$ and $r$, as a function of $r$. Power-law fits are displayed for comparison with mass segregation theory. Bottom panel: the corresponding local value of $\Delta_{AH}(r)$ (equation \ref{eq:DeltaAH}) found for the local values of $N_L(r)$ and $N_H(r)$ shown in the upper panel.}  
    \label{fig:Codetests5e4}
\end{figure}

\begin{figure}
    \centering
    \includegraphics[width=\columnwidth]{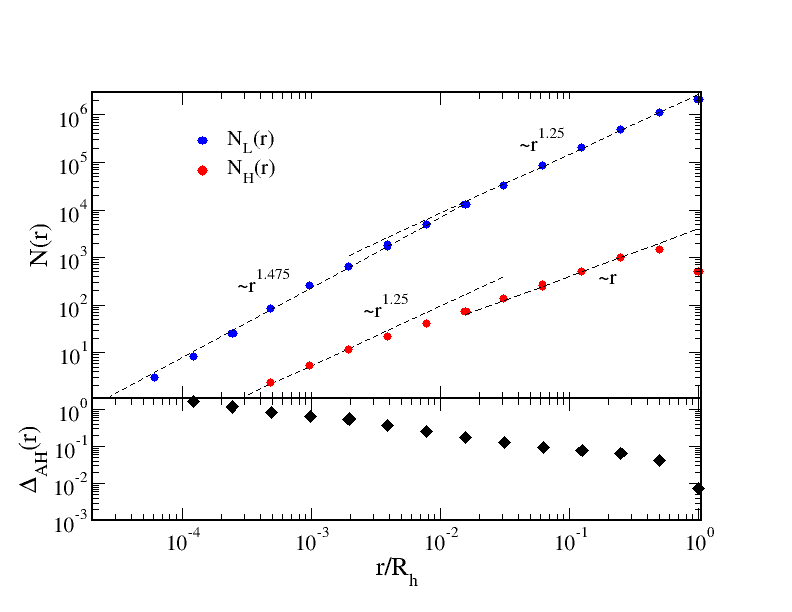}
     \caption{Same as figure \ref{fig:Codetests5e4}, but for the case $\{N_L=3.96\times 10^6,\;N_H=4\times 10^3\}$. Top panel: the steady-state profile at time $\ttb(R_h)$ of the $N_L(r),\;N_H(r)$. Bottom panel: the corresponding local value of $\Delta_{AH}(r)$ (equation \ref{eq:DeltaAH}) found for the local values of $N_L(r)$ and $N_H(r)$ shown in the upper panel.} 
    \label{fig:Codetests4e3}
\end{figure}

Figures \ref{fig:Codetests5e4} and \ref{fig:Codetests4e3} present the results of two simplified simulations which serve as tests of the code, as well as a basic demonstration of mass segregation. The values of $N(r)$ are averaged over the last $10\%$ of the simulation time, which is one cluster relaxation time, $\ttb(R_h)$. In these simulations (and all that follow) the steady-state is typically reached after several tenths of $\ttb(R_h)$, so the assumption of steady state appears to be robust. The simulations were carried out without relativistic corrections (the GW loss terms) and ignoring the possibility of collisions. Stars from disrupted binaries are also not taken into account. The figures show, respectively, the steady state stellar profiles found for the two combinations $\left\{N_L=3.5\times 10^6;\;N_H=5\times 10^4\right\}$ and $\left\{N_L=3.96\times 10^6;\;N_H=4\times 10^3\right\}$. The profiles are displayed in terms of the number of stars in the $i$-th radial shell, with  $r_i$ as the outer radius of that shell ($N(r_i)$ is the number of stars between $r_i/2$ and $r_i$).


In both cases the code reproduces the general trends expected from theory, and also the standard results found in other works, which are mostly numerical solutions of the diffusion equation in $E-J$ space \citep{Merritt2010,BarOretal2013,ZhangAS2023} and $N$-body simulations with a smaller numbers of stars \citep{FreitagASKalogera2006,ASPreto2011,Antonini2014,BASS2018,Panamarevetal2019}. For $N_H=5\times 10^4$ the stellar-mass black holes dominate the two-body scatterings over almost the entire range of the simulation, constituting a weak segregation scenario. Interior to about $0.2R_h$ the steady-state radial profile of the MS stars is numerically consistent with $N_L(r)\propto r^{1.475}$, which is the theoretical estimate of $\alpha_L=3/2+m_L/4m_H$ \citep{BW77} for weak segregation when the heavier objects dominate. It is only in the outer $0.2R_h<r$ region that the lighter stars dominate their own scatterings and their profile turns over to the value typical of strong segregation, with $\alpha_L\approx 7/4$. These trends coincide with the results for the black holes, which interior to about $0.05R_h$ develop a profile that is consistent with the standard BW power law of $N_H(r)\propto r^{5/4}$ $(\alpha_H=7/4)$, while beyond this radius the density profile becomes steeper, as is appropriate for a transition to the strong segregation regime.

The general trends in the case $N_H=4\times 10^3$ are similar, except that the transition between strong and weak segregation occurs around $0.005-0.01R_h$. Again, in the outer part of the cluster the MS stars dominate the scatterings, and so follow a radial profile that is consistent with the BW value of $\alpha_L\approx 1.75$, while the black holes follow a steeper profile, roughly consistent with $\alpha_H=2$. The inner part of the cluster is again consistent with the weak segregation theoretical values of $\alpha_L=1.475$ and $\alpha_H=1.75$. 

While this transition from weak to strong segregation at increasing radii is not the main topic of the current work, it is noteworthy that that it is an inevitable consequence of segregation. As pointed out recently by \citet{LinialSari2022}, even for $N_H<<N_L$, it is always expected that at sufficiently small radii the heavier stars become abundant enough to dominate gravitational scattering, so that in the inner region of the cusp segregation is weak, rather than strong.

This transition from strong to weak segregation in any given profile leads to some further insight. Following \citet{AlexanderHopman2009}, it is customary to evaluate the regime of segregation through the parameter $\Delta_{AH}$
\begin{equation}\label{eq:DeltaAH}
\Delta_{AH}=\frac{N_H m^2_H}{N_L m^2_L}\frac{4}{3+m_H/m_L}\;.
\end{equation}
Values of $\Delta_{AH}>1$ correspond to weak segregation, while $\Delta_{AH}<1$ implies strong segregation. 

Most studies use the global values of $N_L$ and $N_H$ in the cluster to describe the entire system with a single value of $\Delta_{AH}$. Numerical results are then verified by extracting some characteristic value of the powers $\alpha_L$ and $\alpha_H$ (typically at some fiducial radius or by averaging over some finite range of radii). However, by nature segregation causes the value of $\Delta_{AH}$ to change across the cluster, increasing inward. This is demonstrated in the lower panels in figures \ref{fig:Codetests5e4} and \ref{fig:Codetests4e3}, which show the {\it local} values of $\Delta_{AH}(r)$ as a function of radius. Note that in both figures the transition between strong and weak segregation occurs where $\Delta_{AH}$ has a value of a few tenths, as is to be expected based on the relation between $\Delta_{AH}$ and the relevant gravitational times scales \citep{AlexanderHopman2009,PretoAS2010}.

Finally, another noteworthy result is that in both simulations $N_H(r)$ tends to turn over and actually {\it decrease} in the very outer radial shell (between $R_h/2$ and $R_h$), even though this shell is the most voluminous. This result is qualitatively consistent with another prediction by \citet{LinialSari2022}, that heavier stars should be exponentially rare at large radii. However, since the Monte Carlo code described here focuses on occurrences near the SMBH, it is ill-equipped in terms of resolution to quantitatively reproduce this exponential decline.

\section{TDEs, DCs and EMRIs in a Segregated Cluster} \label{sec:MainResults}

We may now turn to consider the main topic of the current work, which is the possible impact of segregation on destructive events near an SMBH. By selectively adding GW losses and high-velocity stellar collisions, the rates of these events can be estimated as a function of the stellar composition of the cluster, namely  as a function of the combination $\{N_L,N_H\}$.

\subsection{Stellar Profiles} \label{subsec:fullprofiles}

\begin{figure}
    \centering
    \includegraphics[width=\columnwidth]{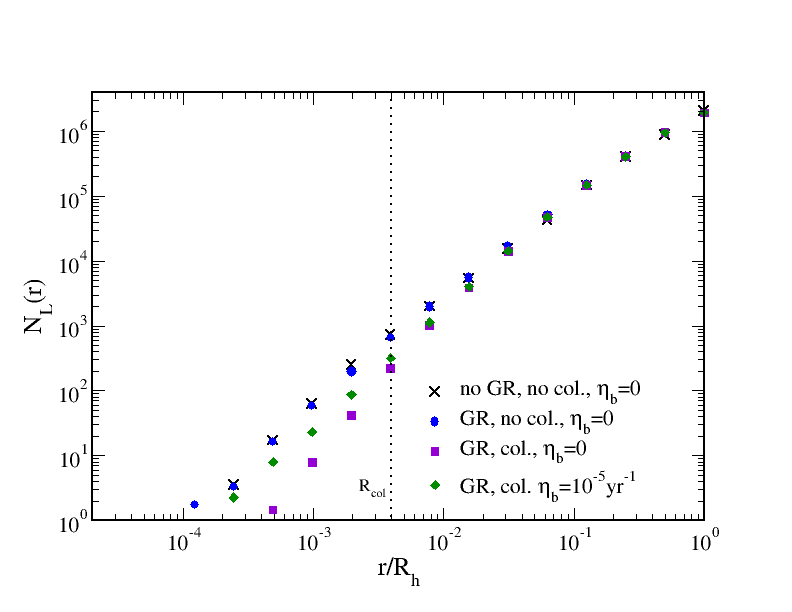}
    \includegraphics[width=\columnwidth]
    {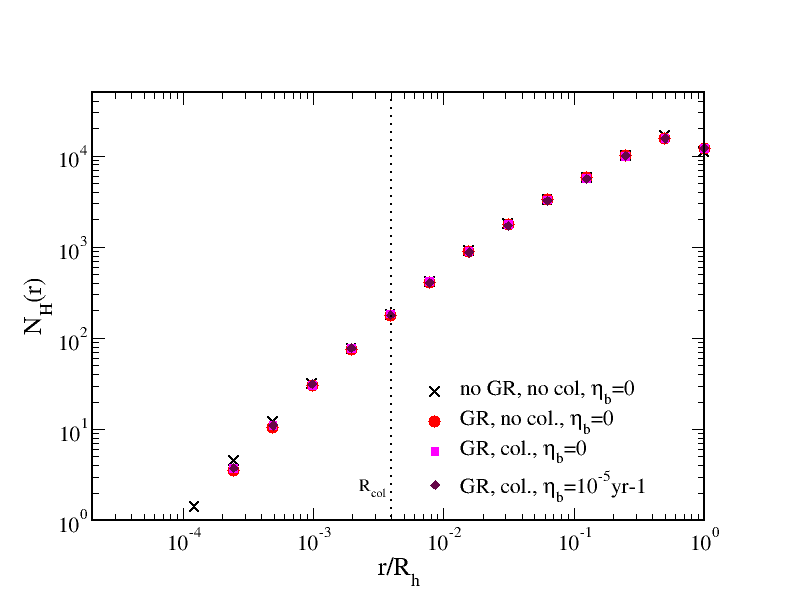}
     \caption{Top panel: the steady-state profile at time $\ttb(R_h)$ of the $m=1\Msun$ MS stars for the case $\{N_L=3.5\times 10^6,\;N_H=5\times 10^4\}$. Shown is the number of stars in each radial shell, $N_L(r)$. GW losses are always included, and the different cases are no collisions (blue circles), collisions but no injected stars (violet squares), and collisions along with injected stars at $\eta_b=10^{-5}\;\invyr$ (green diamonds). Bottom panel: the corresponding profiles of the $10\;\Msun$ stars for no collisions (red circles), collisions but no injected stars (magenta squares), and collisions along with injected stars at $\eta_b=10^{-5}\;\invyr$ (maroon diamonds). Also shown for reference are the results for the code test (no GW and no collisions) from figure \ref{fig:Codetests5e4} (black X's).} 
    \label{fig:profs5e4}
\end{figure}

\begin{figure}
    \centering
    \includegraphics[width=\columnwidth]
    {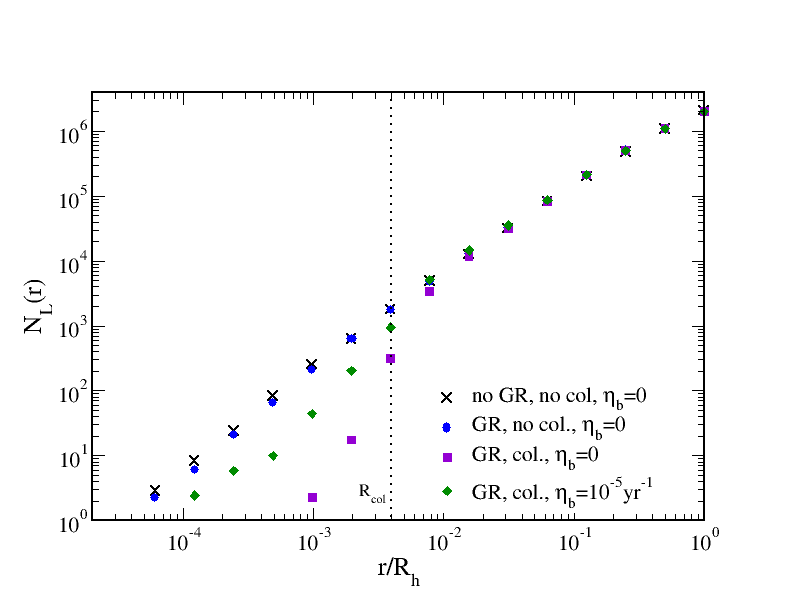}
    \includegraphics[width=\columnwidth]{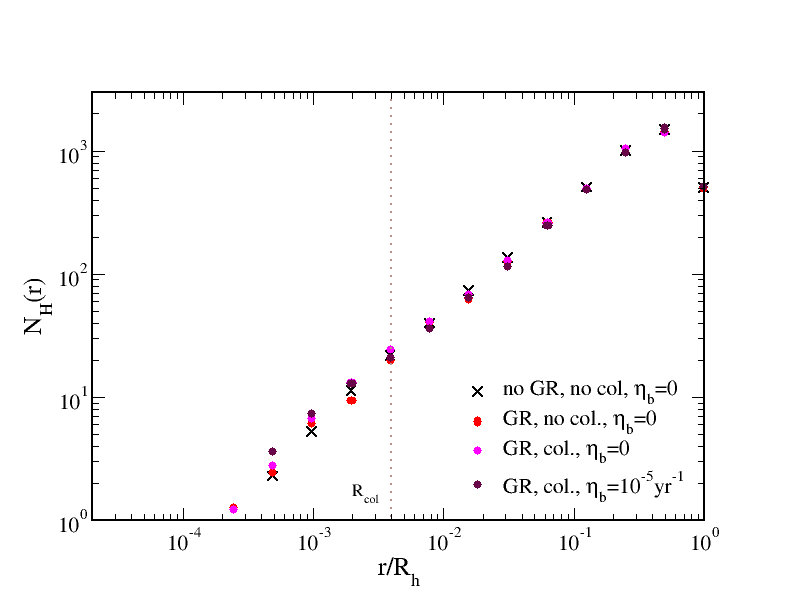}
     \caption{Similar to figure \ref{fig:profs5e4} but for the case $\{N_L=3.96\times 10^6,\;N_H=4\times 10^3\}$. Top panel: the steady-state profile at time $\ttb(R_h)$ of the $m=1\Msun$ MS stars. Bottom panel: the corresponding steady-state profile of the $m=10\Msun$ black holes. Also shown for reference are the results for the code test (no GW and no collisions) from figure \ref{fig:Codetests4e3} (black X's).} 
    \label{fig:profs4e3}
\end{figure}

Further simulations with $N_H=5\times 10^4$ and $N_H=4\times 10^3$, were carried out while including (i) GW losses, and (ii) GW losses and destruction of stars in DCs. 
Figures \ref{fig:profs5e4} and \ref{fig:profs4e3} show the steady-state radial profiles of the two combinations of stars in the cluster, respectively. Again shown are the profiles found after about one cluster relaxation time, $\ttb(R_h)$, well after a steady state has been reached. The results for the test case of no GW losses, no collisions, and no injected stars (figures \ref{fig:Codetests5e4} and \ref{fig:Codetests4e3}) are also shown (black crosses). The top panel in both figures shows the profiles of the MS stars, and the bottom panels that of the black holes.

The inclusion of GW losses alone (no collisions) has a minimal effect on the radial profile. As is to be expected, GW losses play substantial role only close to the SMBH, where the general relativistic timescale is shorter than the gravitational timescales. This causes a minor depletion of stars of both types close the SMBH, so that the profiles deviate from the standard power laws expected from gravitational scatterings alone. As a result, the profiles are very similar to the test cases shown in \S \ref{sec:Simple}. In fact, even those results also had a slight depletion of stars at inner radii due to the finite inner radius stars can sample ($R_T$ or $4R_S$), which causes a deviation from a truly isotropic velocity distribution.

Allowing for DCs induces a dramatic effect on the profile of MS stars in the region interior to $R_{col}$. The collisional time scale for a MS star with orbital parameters $\{r,r_p\}$ can be estimated by \citep{SariFragione2019,BalbergYassur2023}
\begin{equation}\label{eq:Tcol}
T_{col}=\frac{r^2_p}{N_L(r_p)}\left(f_R R_\star\right)^{-2} P(r)\;,
\end{equation}
which for $r<R_{col}$ is shorter than all gravitational time scales in the steady-state profiles derived above. Consequently, regardless of segregation, the region interior to $R_{col}$ becomes practically devoid of MS stars. Scattering from $r>R_{col}$ is too slow and cannot resupply new stars at a sufficient rate to reproduce a profile typical of the collision-free case.

Note, however, that there exists a quantitative difference between the profiles of the two cases $N_H=4\times 10^3$ and $N_H=5\times 10^4$. The former is practically devoid of MS stars, and so is very similar to the single-mass (no black holes) profile. On the other hand, a sufficiently large number of black holes drives the MS stars to exhibit a shallower slope, implying that they spread further inward in the cluster. The qualitative reason is that the abundant heavier black holes decrease the relaxation timescale (equation \ref{eq:T2B}) of the MS stars at $r\gtrsim R_{col}$. In spite of this diffusion being biased (the net drift of dynamical friction drives the MS stars outward), the enhanced diffusion coefficient allows some MS stars to diffuse inward prior to experiencing a collision. A large number of heavier stars thus leads to a finite steady-state penetration of MS stars to smaller radii.

Finally, note that allowing for collisions has practically no effect on the distribution of the black holes. This is to be expected, since at radii interior to $R_{col}$ the black holes dominate the scatterings in the weak segregation regime. Further depletion of the MS stars through collisions is inconsequential. 

\subsection{The Effects of Disrupted Binaries} \label{subsec:AddBin}

Injection of MS stars represents the result of binary disruptions near the SMBH (it is assumed that only MS-MS binaries are relevant for disruptions). The procedure for injecting stars is as follows. For every MS star that is destroyed in any of the channels (TDE, EMRI, or DC), a random number, $\xi$ is drawn and compared to a predetermined probability $0\leq p_b<1$. If $\xi>p_b$ a new star is placed in the outer radial shell $(R_h/2<r<R_h)$ with its orbital parameters drawn randomly assuming a BW profile and an isotropic velocity distribution. For $\xi<p_b$ the star is replaced by an injected star, whose properties are drawn based on additional assumptions regarding the original binary population. In the results shown here it is assumed that the binary separation distance, $a_b$, follows a log-normal distribution: \begin{equation}\label{eq:abspread}
f(a_b)\propto \frac{1}{a_b}
\end{equation}
which conforms with the observed population of binaries \citep{DucheneKraus2013}. This distribution is assumed to lie between two values, $a_{min}=0.01\;$au, and $a_{max}$=$1\;$au. The choice of $a_{max}$ is somewhat arbitrary (although it has to be a relatively small value, since only tight binaries can survive in the dense stellar cluster). Further detail and sensitivity checks can be found in \cite{BalbergYassur2023}. 

Once the original value of $a_b$ has been determined, the injected star is assumed to be captured in an orbit with a periapse $r_p(a_b)$ and an sma $r(a_b)$ of
\begin{equation} \label{eq:R_TB}
r_p(a_b)\approx R_T(a_b)\approx \left(\frac{\MBH}{m}\right)^{1/3}a_b \;\;,\;\; r(a_b)=\left(\frac{\MBH}{m}\right)^{2/3}a_b\;,
\end{equation}
where $R_T(a_b)$ is the distance from the SMBH where binary tidal disruption is expected.
The injected orbit is thus very eccentric, with $e \approx 0.99$. These estimates are consistent with the inferred physics of binary disruption (see, e.g. \citet{Bromleyetal2006,PeretsSubr2012,Rossietal2014}), in which the three-body process with the SMBH often results in the partner being ejected with a large energy. This inference is supported by the existence of hypervelocity stars (HVS) which are observed leaving the plane of the Milky Way with velocities exceeding $1000\;{\rm km\; s^{-1}}$ \citep{Brown2015}; binary disruption is the favored production mechanism of these stars.

Figures \ref{fig:profs5e4} and \ref{fig:profs4e3}
also show the radial profiles of the stars when the parameter $p_b$ is adjusted in each case to produce an assumed injection rate of $\eta_b=10^{-5}\invyr$. GW losses and collisions are also included. 

As is to be expected, binary disruption creates a significant source of MS stars for the inner region of the cluster. Correspondingly, the abundance of MS stars is obviously greater than in the corresponding case with no injected stars ($\eta_b=0$). Since the shortest distances dominate the collision rate, the highly eccentric orbits of the captured stars lead to much shorter collision time scales (equation \ref{eq:Tcol}) than stars which diffuse inward at $R_{col}$. As a result, most injected stars undergo collisions, as is the case for a single-mass population \citep{BalbergYassur2023}. However, some minor differences do arise between the two cases, $N_H=4\times 10^3$ and $N_H=5\times 10^4$. While the collisions dominate, the two-body relaxation and dynamical friction time scales are not infinitely  longer, so to some extent this effect modifies the distribution and orbits of the injected stars. As discussed above, this effect is more pronounced for a larger black hole population, causing the injected stars to sustain a shallower density profile in the inner part of the cluster. These effects carry on into the rates of destructive events that occur near the SMBH, which is the topic of the following subsection.

\subsection{Destructive Events near the SMBH and the Effects of Segregation}

All three destructive channels of MS stars - TDEs, EMRIs, and DCs - affect the steady-state profile of the stellar cluster. Since all three processes occur close to the SMBH, there is some codependence between them and gravitational scatterings, and specifically consequences of mass segregation.

\begin{figure}
    \centering
    \includegraphics[width=\columnwidth]{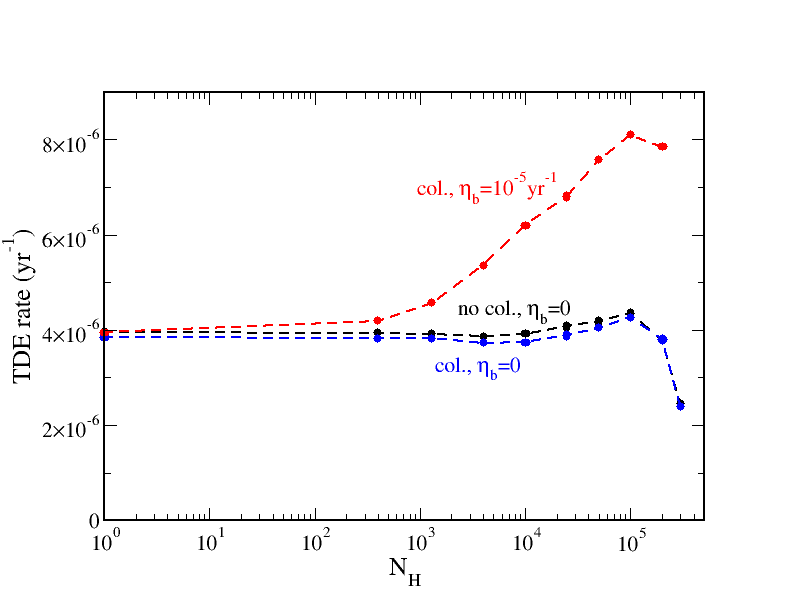}
     \caption{The TDE rate averaged over the last $10\%$ of the simulation time, as a function of the number of $m=10\;\Msun$ black holes, $N_H$. Shown are the results for the simulation with GW losses but no collisions and no injected stars (black), GW losses and collisions but no injected stars (blue), and GW losses and collisions and stars injected at a rate $\eta_b=10^{-5}\invyr$ (red).} 
    \label{fig:TDErates}
\end{figure}

Figure \ref{fig:TDErates} shows the results of the simulations in terms of the TDE rates in the steady state, as a function of the total number of stellar-mass black holes, $N_H$ (when in all cases $\sum m=\MBH$). The curves in the figure correspond to the physical scenarios considered above: no collisions, allowing for collisions, and allowing for collisions when stars are injected (representing disrupted binaries) at a rate of $\eta_b=10^{-5}\invyr$. GW losses are included in all simulations.   

Recall that TDEs are the result of stars being gravitationally scattered into eccentric orbits which end in them plunging to and below $R_T$. In a gravitationally relaxed cluster, this rate is dominated by stars with an sma close to $R_h$, which are the vast majority of the MS stars. The TDE rate can therefore be roughly approximated by \citet{SariFragione2019}
\begin{equation}\label{eq:RTDE_approx}
\mathcal{R}_{TDE}\approx \frac{N_L(R_h)}{\ttb(R_h)\ln(J_C(R_h)/J_{LC})}\;,
\end{equation}
where $J_C(R_h)$ is the angular momentum of a circular orbit at $R_h$, and $J_{LC}=\sqrt{2G\MBH R_T}$ is the "loss cone" value in angular momentum space, which corresponds to  an eccentric orbit with $r_p=R_T$.  For the Milky Way SMBH and a cluster of Sun-like stars, this value is $\mathcal{R}_{TDE}\approx 7\times 10^{-6}\invyr$, which is consistent with the result here, particularly for $N_H=0$. Consequently, mass segregation has a relatively minor effect on the TDE rate, since the black holes are subdominant close to $R_h$. This is indeed reflected in the fact that the TDE rate is almost independent of the existence and number of black holes. The very minor dependence that does exist can be assessed by the rough approximation in equation \ref{eq:RTDE_approx}: As $N_H$ is increased in the simulations, $N_L$ decreases linearly (since the total mass in the cluster is kept fixed), but the two-body relaxation time scale becomes shorter with a quadratic dependence on $N_H$. The latter effect is weak because of the exponential-like decline in the number of heavier stars close to $R_h$, but it still dominates over the former, linear scaling of $N_L$. The TDE rate thus slightly increases for larger values of $N_H$. The opposite is true only for very large values of $N_H$, when $N_L\leq 2\times 10^6$.

The effect of mass segregation is far more pronounced when stars from disrupted binaries are injected at the relatively high rate of $\eta_b=10^{-5}\invyr$. In the case of a single-mass population $(N_H=0)$, the relaxation time experienced by the injected stars is significantly longer than the collision time and so practically all injected stars are destroyed in DCs \citep{BalbergYassur2023}. Hence, they hardly contribute to the TDE rate, which remains unchanged (essentially driven only by stars at $R_h$, as mentioned above). However, in a two-mass system the segregated heavier stars are overconcentrated in the inner part of the cluster (where the stars are injected), and the two-body relaxation time there is clearly shortened, and some injected stars evolve their orbits. Although on average the injected stars gain angular momentum (due to dynamical friction), the stochastic nature of two-body scatterings allows some stars to lose angular momentum so that their periapse drops below $R_T$ and they are destroyed as TDEs. The net result is a sizable increase in the total TDE rate, and for large values of $N_H$ it reaches a factor of $2$ with respect to the single-mass case.

\begin{figure}
    \centering
    \includegraphics[width=\columnwidth]{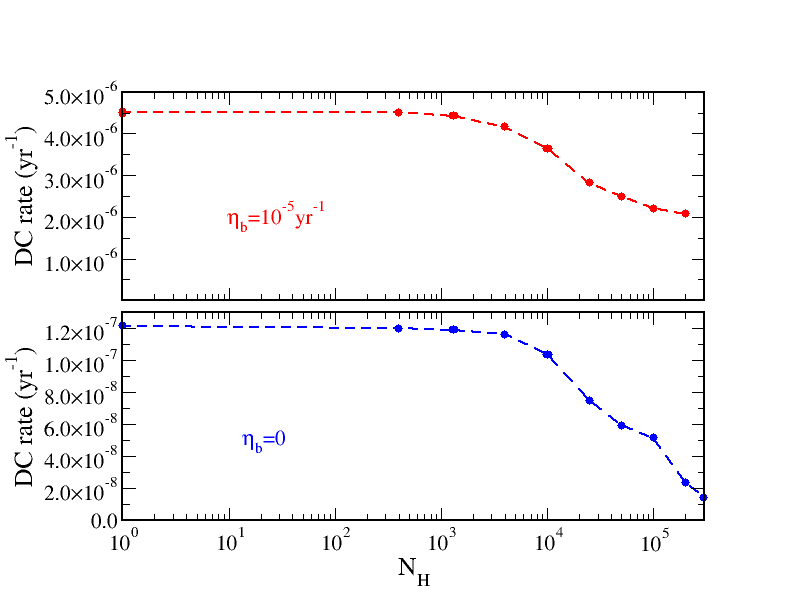}
     \caption{The DC rate averaged over the last $10\%$ of the simulation time, as a function of the number of $m=10\;\Msun$ black holes, $N_H$. Top panel: the rates for the simulations with GW losses and collisions and stars injected at a rate $\eta_b=10^{-5}\invyr$ (red). Bottom panel: the same with GW losses and collisions but no injected stars (blue).} 
    \label{fig:DCrates}
\end{figure}

Figure \ref{fig:DCrates} shows the DC rates in the steady state in the simulation as a function of $N_H$. The results are for either collisions but no injected stars, or collisions and stars injected at a rate of $\eta_b=10^{-5}\invyr$.  

In the absence of segregation (no stellar-mass black holes) the collision time of the MS stars at $R_{col}$ is much shorter than the two-body relaxation time. Since the region interior to $R_{col}$ is essentially depleted (figures \ref{fig:profs5e4} and \ref{fig:profs4e3}), the point of reference in terms of the steady-state DC rate in a cluster without injected stars is approximately \citep{BalbergYassur2023}
\begin{equation}\label{eq:RDC_approx}
\mathcal{R}_{DC}(N_H=0)\approx \frac{1}{2}\frac{N^2_L(R_{col})}{R^3_{col}}v(R_{col})\pi (f_R R_\star)^2\;,
\end{equation}
where $v(R_{col})\sim \sqrt{G\MBH/R_{col}}$ is the typical velocity at $R_{col}$. For the Milky Way parameters and $N_L(N_H=0)=4\times 10^6$, this corresponds to $\mathcal{R}_{DC}\sim 10^{-7}\invyr$, which is indeed the result found in the simulations. 

As already demonstrated above, when heavier black holes are included in the cluster their primary effect  is to redistribute the lighter stars. Since in the steady-state profile the MS stars tend to larger radii, their density at $R_{col}$ is smaller than in a single-mass profile, and the total DC rate decreases appropriately. This effect is marginal for $N_H\lesssim 5\times 10^3$, but becomes significant at higher values. Specifically, note that the suppression of the DC rate exceeds the simple effect of $N_L$ being smaller than $4 \times 10^6$ in the single-mass case. For example, for $N_H=2.5\times 10^4$, the collision rate has dropped to about $0.6\mathcal{R}_{DC}(N_H=0)$, whereas the total number of MS stars in the simulation is reduced by a factor of only $\sim 0.875$ (and $N^2_L$ is smaller by a factor of $\sim 0.765$). For very high values of  $N_H$ (which are probably not realistic), DCs are almost nonexistent.

The DC statistics are substantially different when  stars from disrupted binaries are taken into account. Firstly, the reference DC rate is high, with $\mathcal{R}_{DC}(N_H=0)\lesssim \eta_b/2=5\times 10^{-6} \invyr$. Again, this is simply because the injected stars are almost exclusively destroyed by collisions, with one collision for every two injected stars. Secondly, while the impact of segregation on the total collision rate is also quite significant (for high values of $N_H$), the origin of this effect is different than in the case of the relaxed steady-state profile discussed above. Segregation does not affect the disrupted binaries as the source of candidate stars for collisions (with $\eta_b$ being an external parameter). The main effect here is that a shorter relaxation time allows injected stars to evolve their orbits prior to undergoing a collision. This is a mirror image of the effect that we have already seen in figure \ref{fig:TDErates}: A much greater fraction of the injected stars is destroyed in TDEs, at the expense of DCs. While the total DC rate is still substantial enough to be potentially observable, the suppression induced by mass segregation with respect to $\mathcal{R}_{DC}(N_H=0)$ can be important for $N_H\gtrsim 5\times 10^3$, and reaches $\sim 50\%$ at the highest values.

\begin{figure}
    \centering
    \includegraphics[width=\columnwidth]{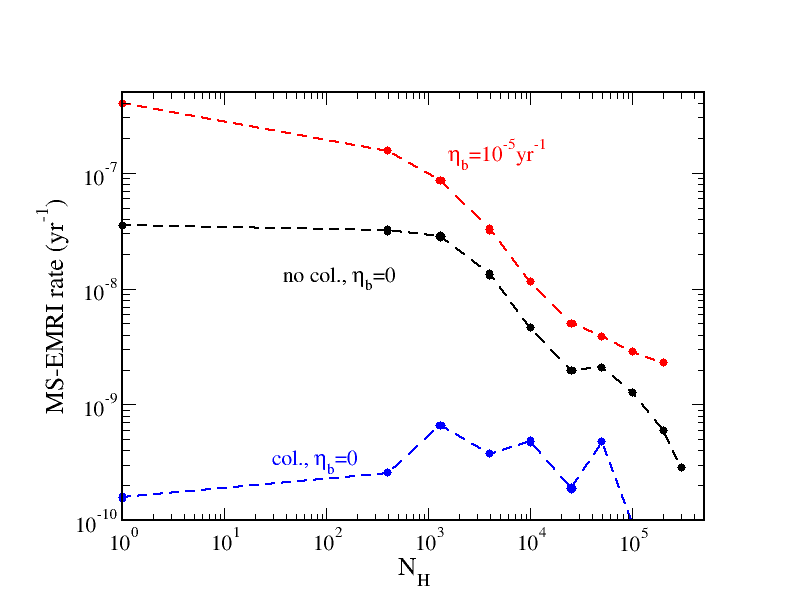}
     \caption{The MS-EMRI rate averaged over the last $10\%$ of the simulation time, as a function of the number of $m=10\;\Msun$ black holes, $N_H$. Shown are the rates for the simulation with GW losses but no collisions (black), GW losses and collisions but no injected stars (blue), GW losses, collisions with stars injected at a rate $\eta_b=10^{-5}\invyr$ (red).} 
    \label{fig:EMRIrates}
\end{figure}

The MS-EMRI channel also terminates with stars being tidally disrupted at $R_T$, but after a "gradual" descent, driven mostly by GW losses. Figure \ref{fig:EMRIrates} shows the MS-EMRI rates calculated in the simulations as a function of $N_H$, again for the cases of no collisions, collisions but no injected stars, and collisions along with stars injected at a rate of $\eta_b=10^{-5}\invyr$.  

By construction, MS-EMRIs are a minor destruction channel in a Milky-Way like cluster. A star is likely to undergo an EMRI only when its GW loss time 
\citep{Peters1964,HopmanAlexander2005}
\begin{equation}
T_{GW}(r,r_p)=\frac{R_S}{c}\frac{\MBH}{m} \left(\frac{r_p}{R_S} \right)^4 \left(\frac{r}{r_p} \right)^{1/2}\ ,
\label{eqn:gw}
\end{equation}
is the shortest of all evolutionary time scales. As shown by \citet{SariFragione2019}, in the absence of collisions, in a single-mass cluster $T_{GW}$ is shorter than $\ttbj$ only in a relatively narrow region in $(r,r_p)$ space, leading to a typical MS-EMRI rate of 
\begin{equation}\label{eq:REMRI_approx}
\mathcal{R}_{EMRI}(N_H=0)\approx \left(\frac{R_S}{R_T}\right)^2 \mathcal{R}_{TDE}\;,
\end{equation}
or about $1\%$ of the TDE rate for Milky Way parameters. This is consistent with result $\mathcal{R}_{EMRI}(N_H=0)\approx 4\times 10^{-8}\invyr$ found for this case in the simulations.

 It is evident that in the simulations without collisions mass segregation has an inhibiting effect on MS-EMRIs. This is a direct result of the dynamical friction aspect of segregation. The presence of black holes close to the SMBH opposes the GW loss effect on the orbital evolution of the lighter MS stars. A similar trend was naturally found by \citet{AharonPerets2016}, that heavier black holes reduce the EMRI rate of other, lower-mass compact objects. Unlike TDEs, which occur through a small number random scatterings, an MS-EMRI progresses through multiple, mostly quasi-spherical orbits. This becomes very unlikely when the drift drives a net energy gain through dynamical friction which counters the energy loss due to GW emission. Systematic inspiral toward $R_T$ is thus suppressed, by an order of magnitude for $N_H\gtrsim 10^4$ and increasingly so (with some numerical scatter) at larger values.

As demonstrated in \citet{BalbergYassur2023}, collisions also suppress MS-EMRIs, since the probability of a star systematically inspiraling all the way from $R_{col}$ to $R_T$ before undergoing a collision is very small. Combined with the aforementioned effect of mass segregation, including collisions in simulations of a relaxed cluster drives the MS-EMRI rate to the level of the computational noise, and is essentially zero.

The MS-EMRI rate is resurrected if there exists a significant source of stars from disrupted binaries. These stars are injected relatively close to the SMBH, where $T_{GW}$ is comparable with the other timescales. If DCs are ignored, the MS-EMRI rate in such a scenario can rise to as much as a few $10^{-6}\invyr$, on the same order of magnitude as TDEs \citep{SariFragione2019}. Once DCs are taken into account, this is no longer the case (since $T_{col}<T_{GW}$, but in the absence of the heavier black holes the MS-EMRI rate can still be as high as a few $10^{-7}\invyr$ \citep{BalbergYassur2023}). It is clear in figure \ref{fig:EMRIrates} that a sufficiently large population of heavier black holes reduces this value significantly, again due to the opposing drift in energy induced by dynamical friction, which counters GW losses. It appears that even if the injection rate is high, the combined effect of DCs and mass segregation (assuming a nontrivial component of black holes) will render MS-EMRIs to be very rare.

\subsection{The Spatial Distribution of DCs}

The observational potential of DCs strongly depends not only on the total rate, but also on their spatial distribution. Collisions at $R_{col}$ will have a typical energy of $\sim G\Msun^2/\Rsun \sim 10^{49}\;$erg, which could obviously power only a weak, short flare. On the other hand, collisions much closer to the SMBH could generate a light curve on par with standard supernovae in terms of the total energy, and perhaps similar to TDEs in terms of some of the observables \citep{BalbergSariLoeb2013,AmaroSeoane2023}. 

\begin{figure}
    \centering
    \includegraphics[width=\columnwidth]{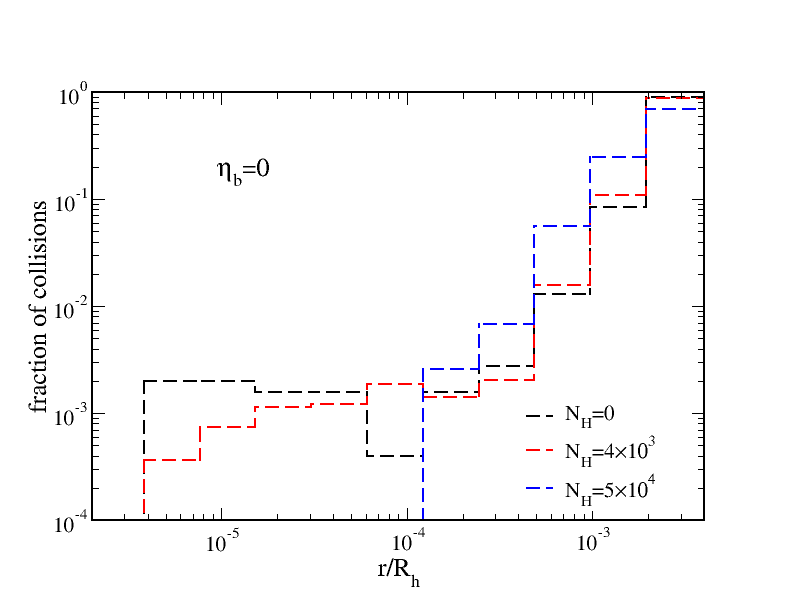}
     \caption{The spatial distribution of the DCs averaged over the second half of the simulation time, for simulations with GW losses, collisions, but no injected stars. Shown is the fraction of collisions in each radial bin (between $r/2$ and $r$) as a function of $r$, for stellar populations with three values of $N_H$ (the number of the $m=10\;\Msun$ black holes). These are $N_H=0$ (black; a single-mass population), $N_H=4\times 10^3$ (red), and $N_H=5\times 10^4$ (blue). }
    \label{fig:COLgridnobin}
\end{figure}

Figure \ref{fig:COLgridnobin} shows the fraction of collisions which occur in the different radial bins in the simulations with no injected stars from binaries. The effects of mass segregation are reflected in the comparison between the three cases of $N_H=0$ (a single-mass population), $N_H=4\times 10^3$ and $N_H=5\times 10^4$. 

Recall that if there are no injected stars, the only source of stars for collisions are those which diffuse inward the cluster from $r>R_{col}$. As long as the relaxation time at $R_{col}$ is longer than the collision time there, collisions tend to be confined close to $R_{col}$, and this is the inferred distribution of DCs shown in in figure \ref{fig:COLgridnobin}. Mass segregation and the presence of black holes does shorten $\ttb(R_{col})$ to some extent, as mentioned above, and so some penetration of MS stars to $r<R_{col}$ is possible. The fraction of collisions that occur in radii $r< R_{col}$ thus increases. This is especially evident for the larger abundance of black holes, $N_H=5\times 10^4$. For example, the fraction of DCs which occur in the radial shell of $5\times 10^{-4} R_h<r<10^{-3} R_h$ (1-2 mpc) is 4 times larger for $N_H=5\times 10^4$ than for $N_H=0$.

Interestingly, this trend is reversed at the smallest radii $(r\lesssim 10^{-4}R_h)$, where for $N_H=0$ and even for $N_H=4\times 10^3$ the fraction of collisions is about $10^{-3}$, whereas for $N_H=5\times 10^4$ it is essentially zero. This is a consequence of the fact already mentioned above in the context of MS-EMRIs, that mass segregation inhibits MS stars from diffusing effectively very deep in the cluster. The net drift caused by dynamical friction prevents stars from reaching very small radii from the SMBH through multiple orbits, dramatically reducing the probability of a collision in this region.

\begin{figure}
    \centering
    \includegraphics[width=\columnwidth]{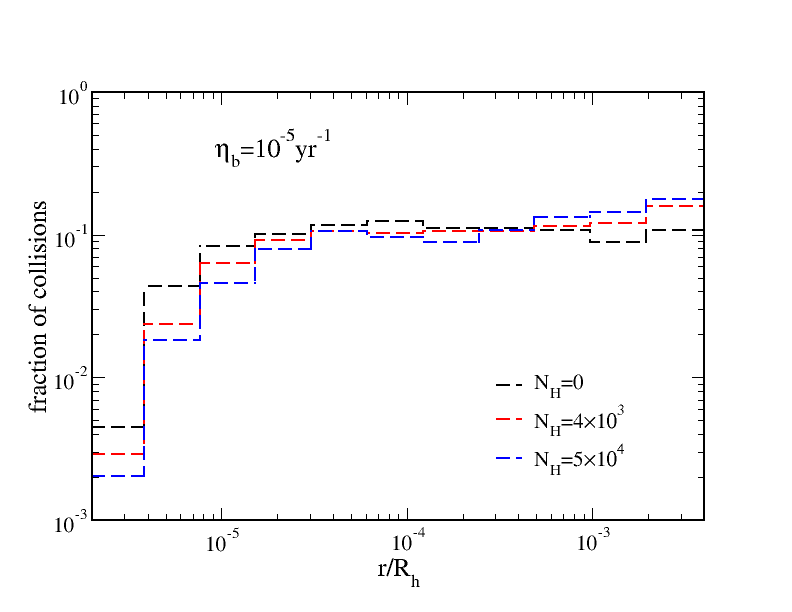}
     \caption{The spatial distribution of the DCs averaged over the second half of the simulation time, for simulations with GW losses, collisions, and injected stars at a rate of $\eta_b=10^{-5}\;\invyr$. The fraction of collisions in each radial bin and the color-coding are the same as in figure \ref{fig:COLgridnobin}.} 
    \label{fig:COLgridbin}
\end{figure}

Some of these dynamics change when there exists a significant source of stars from disrupted binaries. Figure \ref{fig:COLgridbin} shows the fraction of collisions which occur in the different radial bins in the simulations when stars were injected at a rate of $\eta_b=10^{-5}\invyr$. The effects of mass segregation are again reflected in the comparison between the three cases of $N_H=0$, $N_H=4\times 10^3$ and $N_H=5\times 10^4$.  

When the injection rate is high but the population is limited to a single-mass, the DC distribution is almost uniform in the logarithmic radial bins. This result reflects the assumed log-normal distribution of the separation distances of the disrupted binaries, which carries on to a similar distribution of the orbits of the injected stars (when assuming $a_{max}=1\;$au; see also in \citet{BalbergYassur2023}). It is only the radial bins closest to $R_T$ which are naturally underrepresented in this distribution.

A second component of heavier black holes causes some quantitative, but not qualitative, changes in the spatial distribution of the DCs. First, the fraction of collisions closest to $R_T$ decreases, because some of the injected stars that travel through this region do experience sufficient gravitational scattering.  Recall that the injected stars have eccentric orbits, so they are mostly scattered by black holes at $r\sim 100 R_T$, where their number is not negligible. These captured stars can undergo a TDE before participating in a collision. The implication is that mass segregation not only enhances the TDE rate, but does so mostly at the expense of the most energetic collisions. The fraction of collisions that involves energies greater than $10^{52}$ erg decreases from about $0.15$ in a single-mass population $(N_H=0)$ to about $0.10$ for $N_H=5 \times 10^4$. Note that this effect comes on top of an overall reduction of about $50\%$ in the total collision rate, so clearly mass segregation can constrain the observational potential of the highest-energy DCs. This preliminary conclusion requires more scrutiny, especially in terms of the assumptions regarding $a_{max}$ and the mass distribution of the stars, and also on the mass of the SMBH when applied to other galaxies.

The suppression of the fraction of collisions at the smallest distances from the SMBH is offset by an increased fraction of collisions at the opposite end, close to $R_{col}$. The enhancement of the fraction in this particular range for large $N_H$ is again a result of the drift induced by the dynamical friction aspect of mass segregation, which enhances the partial probability of a captured star to reach larger distances from the SMBH. 

\section{Conclusions and Discussion} \label{sec:Conclusions}

The violent destruction of stars is one aspect of the dynamics that shape the dense cluster of stars which surrounds an SMBH. While the deep potential of the SMBH dominates the instantaneous motions of the stars, and gravitational interactions among the stars drive the long-term evolution of their orbits, destruction preferentially removes stars with certain orbital properties. Destruction thus also impacts the steady-state profile. Moreover, given the large energies involved in stellar destruction, such events allow for unique observable transients, especially in an era of high-cadence surveys, such as ZTF \footnote{https://www.ztf.caltech.edu/} \citep{zTF}, ASAS-SN \footnote{https://www.astronomy.ohio-state.edu/asassn/} \citep{ASASSN}, and the upcoming ULTRASAT \footnote{https://www.weizmann.ac.il/ultrasat/} mission \citep{ULTRASAT}, and advanced GW astronomy, such as the Laser Interferometer Space Antenna \footnote{https://www.lisa.nasa.gov/}, \citep{LISA}.

There exist three potential channels for the destruction of MS stars near an SMBH. Stars may be scattered into extremely eccentric orbits, which will occasionally result in a star being tidally disrupted if it ventures too close to the SMBH (a TDE). Such an event is followed by an optical flare, first through disruption and then as the debris accretes onto the SMBH \citep{Gezari2021}. Stars sufficiently close to the SMBH also evolve their orbits through GW radiation energy loss and might inspiral and disrupt "gradually" as they approach and cross the tidal disruption radius, $R_T$. In this case (an MS-EMRI), a possible electromagnetic flare will be accompanied by GW emission in the millihertz frequency band \citep{Alexander2017}. Finally, a high-velocity collision between two stars may be completely destructive (a DC), initiating a supernova-like event in terms of the energy involved \citep{BalbergSariLoeb2013, AmaroSeoane2023}.

Since all destructive channels require stars to pass through the inner part of the cluster, and possibly evolve their orbital paths in this region, the event rates can be sensitive to the details of the stellar profile close to the SMBH. The focus of this work is on a first attempt to self-consistently study the event rate of the stellar destruction channels, including DCs, in a "mass segregated" profile \citep{BW77,AlexanderHopman2009}. Equipartition of energy in gravitational scatterings should create a segregated overabundance of heavier objects close to the SMBH. Such objects, and specifically black holes with masses of a few solar masses, shorten the relaxation time scales close to the SMBH, and accelerate diffusion in angular momentum and energy of the MS stars. Furthermore, the accelerated diffusion is also biased, with an average drift that increases the lighter stars' angular momentum and decreases their gravitational energy (following the convention that the potential energy of a bound object is positive).  

The analysis is based a simplified numerical Monte Carlo code of stellar dynamics in galactic centers, based on the specific version by \cite{SariFragione2019} and the self-consistent tracking of DCs added by \citet{BalbergYassur2023}. Here the code was extended to allow for a two-mass population which includes $N_L$ and $N_H$ stars of masses $1\;\Msun$ and $10\;\Msun$, respectively (the latter being black holes). Also added were gravitational scattering terms which reproduce the effects of dynamical friction between populations of unequal mass. The study was performed assuming a Milky Way SMBH of $\MBH=4\times 10^6\;\Msun$. The focus here is on the steady-state profiles and rates, which are typically reached within a few tenths of the global cluster relaxation time. The time-dependent evolution of the event rates is not analyzed here (and requires some further assumptions not included in the code), but it is noteworthy that rates of various destructive channels may evolve differently in time (see, e.g., the case of TDEs and EMRIs in \citet{Broggietal2022}). The estimates derived here should therefore be applicable to a Milky Way-like galaxy, whose age is approximately one relaxation time; realistic estimates for shorter time scales, and possibly also for longer ones, require further analysis.

It is constructive to begin with a general observation found here (\S \ref{sec:Simple}) that for almost any value of $N_H$, gravitational scattering in the inner part of the cluster is dominated by the heavier objects. These objects are sufficiently overconcentrated in this region so that the dynamics are in the regime of "weak" segregation \citep{BW77}. This is the case even if the overall values of $N_L$ and $N_H$ correspond to the regime of "strong" segregation \citep{AlexanderHopman2009,PretoAS2010}, in which the lighter objects are much more abundant and dominate the dynamics; see \citet{LinialSari2022} for an analytical interpretation of this result. It is in this context that the trends regarding the rates of destructive events should be evaluated.

TDEs are dominated by MS stars orbiting close to $R_h$ (the SMBH radius of influence), which are scattered by other stars into very eccentric orbits that cross $R_T$. Correspondingly, in a gravitationally relaxed cluster, the TDE rate is almost insensitive to the presence of a second component of black holes (a few $10^{-6}\invyr$ for the Milky Way case; see also in \citet{Panamarevetal2019}) . A minor enhancement of the TDE rate does arise because a small population of black holes near $R_h$ shortens the relaxation time there. On the other hand, the total DC rate in such a cluster does depend on the number of black holes in the population. This is the case because in the steady state of a relaxed cluster DCs deplete the MS stars in the inner region $r<R_{col}$, since the collision time scale there is much shorter than the gravitational time. The presence of a nontrivial number of heavier black holes at $r\gtrsim R_{col}$ in the segregated profile creates a net drift effect, which lowers the rate that MS stars diffuse inward to $R_{col}$, and correspondingly reduces the DC rate by as much as tens of percent for larger values of $N_H$ ($\geq 10^4)$. Finally, the black holes dramatically reduce the number of stars that can inspiral toward the SMBH through a systematic orbital decay driven by GW emission, since it too is countered by the same drift. This effect tends to reduce the MS-EMRI rate even when DCs are ignored in the simulations (by as much as an order of magnitude for large $N_H$), and makes MS-EMRIs essentially nonexistent once DCs are taken into account. 

The qualification of a "gravitationally relaxed cluster" used in the previous paragraph relates to the case in which there is a negligible addition of stars to the inner part of the cluster from binary tidal disruption. Such disruptions "inject" stars into tight, bound orbits close to the SMBH \citep{Hills1988}, with high $(e\approx 0.99)$ eccentricities. These stars are primary candidates for destruction close to the SMBH. When assuming a single-mass population \citep{BalbergSariLoeb2013,BalbergYassur2023}, in steady state, the collision time of these stars is much shorter than any gravitational time, so they are predominantly destroyed in DCs, at a rate of $\sim\eta_b/2$, where $\eta_b$ is the injection rate. However, in a mass segregated cluster, a significant presence of black holes close to the SMBH shortens the gravitational relaxation time experienced by the injected stars, and allows some of them to be scattered into tidally disruptable orbits. As a result, a significant black hole population directs some of the injected stars to the TDE channel, at the expense of DCs. The effect can be as a large as $50\%$ of the DC rate for $N_H>5\times 10^4$. Since some of the stars are injected very close to the SMBH, the MS-EMRI rate is never zero in this scenario, although it still remains susceptible to significant suppression (more than an order of magnitude) in the presence of a large black hole population, again due to the inherent drift outward mentioned above.

While TDEs and MS-EMRIs are unique to the immediate vicinity of $R_T$, DCs can occur essentially anywhere in the range $R_T<r<R_{col}$, which in the Milky Way case spans 3 orders of magnitude in radius. The spatial distribution of collisions is thus important in terms of the corresponding energy, and therefore of the observational potential. As mentioned above, in the absence of injected stars representing disrupted binaries, most collisions occur close to $R_{col}$. When the black hole population is large $(N_H\gtrsim 5\times 10^3)$, they shorten the relaxation time at $R_{col}$ and more MS stars actually penetrate inward to $r\lesssim R_{col}$, increasing the fraction of collisions there (while the total rate of collisions is reduced). On the other hand, for large values of $N_H$ the fraction of the most energetic collisions, those near $R_T$ is zero, rather than $\sim 10^{-3}$ for lower values (and $N_H=0$). The net drift for large values of $N_H$ completely prevents MS stars from evolving their orbits to acquire very small values of $r_p$, thus disallowing the most energetic DCs.

A similar, though only quantitative, effect arises when the rate of injected stars is large. Again the net drift will tend to reduce the fraction of collisions at the smallest radii (note that for eccentric orbits, collisions close to periapse are most likely). For the particular parameters examined here, which include binaries with original separation distances spread in a log-normal distribution between $a_b=0.01\;$au and $a_b=1\;$au, the fraction of the most energetic collisions with energies greater than $10^{52}\;$erg drops from $0.15$ for $N_H=0$ to $0.10$ for $N_H=5\times 10^4$. This reduction is in addition to the total collision rate dropping by about a factor of two for the same range of $N_H$. A wider parameter survey is required, but it seems that mass segregation could reduce the rate of highest-energy (and most observable) collisions by tens of percents, if the typical fraction of heavy objects is significant.

While the main trends found here appear robust, there are certainly some important points for further consideration. First and foremost, mass segregation should be investigated with a full, realistic mass function. The two-mass system used here probably captures the main qualitative trends, but obviously quantitative issues require further scrutiny, especially if the fractions of other compact objects (white dwarfs and neutron stars) are large. Another issue is the possible outcome of a collision between an MS star and and a black hole. At the highest possible velocities the black hole probably passes through the star without destroying it (see the Appendix in \citet{MetzgerStone2017}), but close to $R_{col}$, or even at $r\gtrsim R_{col}$ partial or full destruction may be possible. This detail should also be expanded to examine the possible role of low-velocity collisions of two MS stars in the outer parts of the stellar cusp. These more likely result in mass transfer, mass loss, and mergers \citep{FreitagBenz2005}, but multiple collisions may, however, be destructive. In any case, low-velocity collisions may accumulate to quantitative effects on the dynamics of the cluster \citep{Sillsetal2005,DaleDavies2006,Roseetal2023}. In particular, they might increase the fraction of stellar-mass black holes \citep{Roseetal2022}, and generate a further complexity in the segregated profile. 

 Finally, it must be stressed that the Monte Carlo averaged scheme is only approximate, and does not cover the full range of possible effects. The dense stellar cluster generates coherent torques between slowly precessing orbits, namely resonant relaxation processes \citep{KocsisTremaine2011,KocsisTremaine2015}. In principle, this effect is mostly relevant at radii beyond the DC regime ($r\gtrsim 1000$ au), since  closer to the SMBH relativistic precession decouples individual orbits from the residual torques of the background stars \citep{BarOrAlexander2016}. This assessment should be reevaluated when considering a stellar profile which is modified by injected stars and mass segregation. Also not considered here are short-range, large-angle scatterings. While generally rare, such scatterings may be effective enough to alter the orbital evolution with respect to the prediction based on multiple small-angle scatterings. Both issues mentioned above are better addressed by designated $N$-body simulations, which will be reported in a separate work (G.~Yassur and S.~Balberg, 2023, in preparation). 

\section{Acknowledgements}

I wish to thank Pau Amaro-Seoane, Reem Sari, and Gilad Yassur for very useful discussions and comments. I also thank the anonymous referee for valuable comments and observations.

\bibliography{SegCol.bib}{}
\bibliographystyle{aasjournal}

\end{document}